\documentstyle[12pt]{article}
\relax
\def\gsim{{~\raise.15em\hbox{$>$}\kern-.85em \lower.35em\hbox{$\sim$}~}}
\def\lsim{{~\raise.15em\hbox{$<$}\kern-.85em \lower.35em\hbox{$\sim$}~}}
\begin{document}
\begin{titlepage}
\vfill

\hskip 4in {ISU-HET-99-6}

\hskip 4in {July 1999}
\vspace{1 in}
\begin{center}
{\large \bf Constraints on $s\rightarrow d \gamma$ 
from Radiative Hyperon and Kaon Decays}\\ 

\vspace{1 in}
{\bf  Xiao-Gang~He$^{(a,b)}$} 
{\bf  and G.~Valencia$^{(c)}$}\\
{\it  $^{(a)}$ Department of Physics,
               National Taiwan University,
               Taipei 10674}\\
{\it  $^{(b)}$ School of Physics,
               University of Melbourne,
               Parkville, Vic. 3052}\\
{\it  $^{(c)}$ Department of Physics and Astronomy,
               Iowa State University,
               Ames IA 50011}\\
\vspace{1 in}
\end{center}
\begin{abstract}

The quark-level process $b \rightarrow s \gamma$ has been used 
extensively to place constraints on new interactions.
These same interactions can be further constrained from the enhancement 
they induce in the quark-level $s \rightarrow d \gamma$ transition, 
to the extent that the short distance contributions can be separated 
from the long distance contributions. We parameterize what is known 
about the long distance amplitudes and subtract it from the data in 
radiative hyperon and kaon decays to constrain new interactions.

\end{abstract}

\end{titlepage}

\clearpage

\section{Introduction}

The decay mode $b \rightarrow s\gamma$ has been used to place 
constraints on physics beyond the standard model \cite{bsg}. The mode is 
particularly useful constraining new interactions which remove the 
chirality suppression that occurs in the standard model. In this case the 
amplitude is enhanced by factors of a heavy mass scale relative to the 
$b$ quark mass. The same type of new physics enhances the 
$s \rightarrow d \gamma$ transition by a factor of a heavy mass 
scale relative to the strange quark mass. 
In models in which the enhancement is as large as one can expect 
on dimensional grounds, that is $m_b/m_s \sim 30$, 
it is possible to place interesting constraints on the new physics 
from $s \rightarrow d \gamma$ 
even though the decay modes involved 
are dominated by long distance physics \cite{singer}.

After introducing our notation for the effective interaction 
responsible for the $s \rightarrow d \gamma$ transition, we 
study the physical radiative hyperon 
decay amplitudes and the radiative kaon decays of the form 
$K \rightarrow \pi \pi \gamma$. In both cases we expect   
the amplitudes to be dominated by long distance physics. We 
describe this long distance physics guided by chiral perturbation 
theory and subtract it from the physical amplitudes in order to 
constrain the new, short-distance, interactions.

In the last two sections we 
illustrate two types of models in which the short distance 
transition can be significantly enhanced with respect to the standard 
model\footnote{These modes also enhance the transitions 
$d\rightarrow d^\prime g$ and were discussed in Ref.~\cite{kagan}.}  
We are not interested here in the specific details of the  
models, and, for this reason,  we only consider the effective low energy  
operators that the new models may generate. We illustrate the 
effects of left-right symmetric models \cite{chomi,fuji} and of generalized 
supersymmetric theories \cite{ggms}.

\section{Short distance $s \to d\gamma$ in the standard model}

The low energy effective Hamiltonian responsible for the 
$s\to d \gamma (g)$ transition can be written as

\begin{eqnarray}
H_{eff} = \sqrt{2}G_F {V_{id}^*V_{is} \over 16 \pi^2}
\bar d \biggl[ 
\biggl(g_s c_{11}^i T^aG_a^{\mu\nu} + ec_{12}^i F^{\mu\nu}
\biggr)\sigma_{\mu\nu}
\biggl(m_s P_R +m_d P_L \biggr) \biggr]s ~+~{\rm h.c.},
\label{sdgver}
\end{eqnarray}
where $G^{\mu\nu}_a$ and $F^{\mu\nu}$ are the gluon and photon field 
strength tensors respectively, and $P_{L,R}\equiv (1\mp\gamma_5)/2$. 
In the standard model, (SM), the coefficients $c_{11}^i$ and 
$c_{12}^i$ are given at the one loop level without QCD corrections  
by\cite{bsgamma}

\begin{eqnarray}
c^i_{11} = {x_i(2+5x_i-x_i^2)\over 4(1-x_i)^3} + {3x_i^2\over 2(1-x_i)^4}\ln 
x_i \nonumber \\
c^i_{12} = {x_i (7-5x_i - 8x_i^2)\over 12(1-x_i)^3} + 
{x_i^2(2-3x_i)\over 2(1-x_i)^4} \ln x_i,
\end{eqnarray}
where $x_i = m_i^2/m_W^2$. This contribution to $c_{12}$ from charm and up 
quarks is negligibly small. Nevertheless, QCD corrections enhance the 
charm contribution considerably as first discussed in Ref.~\cite{svz}, 
we find $c_7^c \approx 0.13$.

We assume in this paper that any contribution beyond the SM is due 
to heavy degrees of freedom which are integrated out at the $W$ mass 
scale and obtain the coefficients at a hadronic mass scale 
$\mu\sim 1$~GeV using the expressions \cite{bef,running}
\begin{eqnarray}
c_{12}(\mu) &=& c_{12}^{SM}(\mu) + c_{12}^{new}(\mu),\nonumber\\
c^{new}_{12}(\mu) &=& \eta^{16/(33-2n_f)}c^{new}_{12}(m_W)
+{8\over 3} (\eta^{14/(33-2n_f)}-\eta^{16/(33-2n_f)})c^{new}_{11}(m_W),
\label{running}
\end{eqnarray}
where $\eta = \alpha_s(m_W)/\alpha_s(\mu)$, and $n_f$ is the number of active
quarks. 

For our phenomenological discussion of physics beyond the standard 
model we will find it convenient to use instead the effective 
Lagrangian
\begin{equation}
{\cal L} = {e G_F \over 2} c \bar{d}\sigma_{\mu\nu}(1+\gamma_5)s 
F^{\mu\nu},
\label{cdef}
\end{equation}
in terms of which the standard model reads
\begin{equation}
c_{SM} = {\sqrt{2} m_s \over 16 \pi^2} \sum_i V_{id}^* V_{is} c_{12}^i
\label{csm}
\end{equation}
Although we will refer to the coefficient $c$ as if it were unique, 
new interactions may induce this operator with opposite chirality. 
The distinction, however, is irrelevant for our purpose. 
Numerically we will use $\alpha_s(m_Z) = 0.119$,  
the CKM matrix of Ref.~\cite{pdb} in the Wolfenstein parameterization, 
and $m_s = 150$~MeV. This corresponds to $c_{SM} \approx 0.04$.

\section{Radiative Hyperon Decays}

The effective Lagrangian for radiative hyperon decays is usually 
written in the form:
\begin{equation}
{\cal L}(B_i \rightarrow B_f \gamma) = 
{e G_F \over 2} \overline{B}_f\bigl(a + b \gamma_5\bigr)
\sigma^{\mu\nu}B_i F_{\mu\nu}
\label{eflag}
\end{equation}
Each decay mode is then characterized by the constants $a$ and $b$ which 
have both real and imaginary (absorptive) parts. The two observables 
are the decay rate and the asymmetry parameter, which are given in terms 
of $a$, $b$ and the photon energy, $\omega$, by:
\begin{eqnarray}
\Gamma (B_i \rightarrow B_f \gamma(\omega)) &=& 
{G_F^2 e^2 \over \pi}\bigl(|a|^2 + |b|^2\bigr)\omega^3 \nonumber \\
{d\Gamma \over d\cos\theta} &\sim &  1+\alpha \cos\theta \nonumber \\
\alpha &=& {2{\rm Re}(ab^\star)\over |a|^2 + |b|^2}
\label{observables}
\end{eqnarray}

The latest numbers found in the particle data book \cite{pdb} 
are shown in Table~\ref{t; defs}. Some of these measurements will be 
improved by the KTEV experiment at Fermilab.

\begin{table}[htb]
\centering
\begin{tabular}{|c|c|c|c|c|} \hline
  Mode & $\Gamma \times 10^{15}$ MeV & $\omega$ MeV &
  $\sqrt(|a|^2+|b|^2)$ MeV & $\alpha$  \\ \hline 
& & & &  \\
$\Lambda \rightarrow n \gamma$ & $4.38 \pm 0.38$ & 162.22 & 16.07 & - \\
$\Sigma^+ \rightarrow p \gamma$ & $10.13 \pm 0.41$ & 224.59 & 15.00 &
$-0.76 \pm 0.08$ \\
$\Xi^- \rightarrow \Sigma^- \gamma$ & $0.51 \pm 0.09$ & 118.06 & 8.83 & 
$1.0 \pm 1.3$ \\
$\Xi^0 \rightarrow \Lambda \gamma$ & $2.4 \pm 0.4$ & 184.13 & 9.85 &
$0.43 \pm 0.44$   \\
$\Xi^0 \rightarrow \Sigma^0 \gamma$ & $7.94 \pm 0.91$ & 116.57 & 35.54 &
$0.20 \pm 0.32$  \\
& & & & \\ \hline
\end{tabular}
\caption{Radiative hyperon decay data. We present the particle data 
book values for the rate and asymmetry parameters as well as the 
corresponding value for $\sqrt(|a|^2+|b|^2)$.}
\label{t; defs}
\end{table}
The long distance contributions to these decays have been studied 
within the context of chiral perturbation theory in Ref.~\cite{jenk,chang}. 
The authors of Ref.~\cite{jenk} find that the imaginary parts of 
$a$ and $b$ are well known and that 
the real part of $b$ is also known. They also find that the real part 
of $a$ cannot be predicted or even estimated reliably. For this reason 
they treat $a$ as a free parameter and attempt to fit the data. 
In the following table we summarize these results.

\begin{table}[htb]
\centering
\begin{tabular}{|c|c|c|c|c|c|} \hline 
Mode & $a_I$ & $b_R$ & $b_I$ & $a_{SD}(SU(6))$ &
$a_R(fit)$  \\ \hline 
& & & & &  \\
$\Lambda \rightarrow n \gamma$ & $-0.68$ & $11.11\pm 1.2$ & $11.21$ &
$\sqrt{3/2}\; c$ & $2.8\pm 18$ \\
$\Sigma^+ \rightarrow p \gamma$ & $6.18$ & $-1.21$ & $-0.53$ & 
$-1/3\; c$ & $13.6 \pm 9$ \\ 
$\Xi^- \rightarrow \Sigma^- \gamma$ & $-1.55$ & $-7.26$ & $-12.34$ & 
$-5/3 \;c$ &$0^\star$  \\ 
$\Xi^0 \rightarrow \Lambda \gamma$ & $0$ & $-2.47 \pm 2.12$ & $0$ & 
$-1/\sqrt{6}\; c$ &$9.5\pm10$ \\
$\Xi^0 \rightarrow \Sigma^0 \gamma$ & $0$ & $2.52 \pm 1.22$ & $0$ & 
$5/(3\sqrt{2})\; c$ &$35\pm 85$ \\
& & & & &  \\ \hline
\end{tabular}
\caption{Radiative hyperon decay amplitudes as described in the text 
in units of MeV. An 
entrance $0^\star$ indicates that a fit is not possible.}
\label{t; amps}
\end{table}

To construct Table~\ref{t; amps} we have used the imaginary parts of $a$ and 
$b$ from Ref.~\cite{jenk}, which are reliable, and we have also used 
their estimate\footnote{Except that we allow a larger 
uncertainty by doubling the maximum value used for the unknown 
counterterm that occurs in chiral perturbation theory \cite{jenk}.}  
for the real part of $b$.
The column labeled $a_R(fit)$ shows the required value of $a_R$ 
to reproduce the measured rates when combined with the known values 
of $a_I,~b_I$ and $b_R$. It is not possible to fit the measured 
rate for $\Xi^- \rightarrow \Sigma^- \gamma$ in this way. 
In the column labeled $a_{SD}(SU(6))$ we show the short distance 
contribution from the operator of Eq.~\ref{cdef} using $SU(6)$ 
wave functions to compute the hadronic matrix elements. 

In most cases the new physics contribution to $a_R$ will be identical 
to the contribution to $b_R$, so we can bound the coefficient $c$ 
by requiring that $a_{SD}(SU(6))$ be less than $a_R(fit)/\sqrt{2}$. 
The best bound is obtained from the mode $\Lambda \rightarrow n \gamma$ 
and it is
\begin{equation}
|c(\mu)| \lsim 12~{\rm MeV}
\label{hypbound}
\end{equation}
We have not used the asymmetry parameters because the only one that 
is well measured is not understood \cite{jenk}.

\section{Radiative Kaon Decays}

In this section we look at decays of the form $K\rightarrow \pi \pi
\gamma$. We start with the decay $K_L \rightarrow \pi^+ \pi^- \gamma$ 
in which the ``direct emission'' has been measured (bremsstrahlung 
is subtracted from the full amplitude). Assuming CP conservation 
we can write this direct emission amplitude in the form 
\begin{equation}
{\cal M}= i g_8 {2 \sqrt{2} eG_F\lambda f_\pi^2 \over M_K^3}
 \xi_M(z,\nu) \epsilon_{\mu\nu\alpha\beta}p^{+\alpha}
p^{-\beta}k^\nu \epsilon^\mu
\end{equation}
We use the notation of Ref.~\cite{lv}, $\lambda\approx 0.22$ is the 
sine of the Cabibbo angle and $g_8 \approx 5.1$.
$\xi_M$ is a form factor that depends on the photon energy
in the kaon rest frame, $z = E_\gamma /M_K$, and on a pion energy 
difference, $\nu = (E_{\pi^+}-E_{\pi^-})/M_K$. At leading order in 
chiral perturbation theory, it is just a constant which can be fit 
to the data to obtain \cite{lv} 
\begin{eqnarray}
\xi_M &\equiv & {M_K^3 \over 8 \pi^2 f_\pi^3}F_M e^{i(\delta^1_1
-\delta^0_0)}\nonumber \\
|F_M| &=& 0.94 \pm 0.06
\end{eqnarray}
Unfortunately, the analysis of this decay mode is more complicated than 
this. There is experimental evidence for significant variation of 
the form factor with $z$. In particular, if we introduce only a 
slope term into the form factor, 
\begin{equation}
\xi_M \equiv  {M_K^3 \over 8 \pi^2 f_\pi^3}F_M (1+c_M E_\gamma /M_K)
e^{i(\delta^1_1 -\delta^0_0)}
\end{equation}
E731 has extracted a value of $c_M=-1.7 \pm 0.5$ \cite{e731} which 
changes the overall constant to $F_M = 1.49 \pm 0.04$. A careful 
analysis in Ref.~\cite{enp} parameterizes the  long-distance 
contributions to $F_M$ in terms of one constant (their $k_F$) 
finding that a typical range is $0.3 < F_{M,LD} < 0.9$. This 
analysis, however, might change in view of the latest results 
from E799 \cite{dpf}. 

In this case it is not possible to separate long distance and 
short distance contributions to $F_M$. To be conservative, therefore, 
we demand that any new physics contribution to $F_M$ be at most 
equal to the measured $F_M$. As we have seen, there is large 
uncertainty at present in the extraction of $F_M$ which can range 
from 0.3 to 1.5. For definiteness we use
\begin{equation}
\biggl|F_M\biggr|_{\rm New} < 1
\label{fmbound}
\end{equation}

The decay $K^+ \rightarrow \pi^+ \pi^0 \gamma$ also has 
long distance contributions that are not known precisely and 
it is not as well measured as $K_L \rightarrow \pi^+ \pi^- \gamma$. 
The decay $K_L \rightarrow \pi^0 \pi^0 \gamma$ 
has not been seen, and in chiral perturbation theory, starts 
at order $p^6$ \cite{kambor} making a separation of long and short 
distance contributions even harder. For these reasons we do not 
consider additional constraints from these modes.

To place a bound on new physics we need to estimate the 
contribution of the short distance operator of Eq.~\ref{sdgver} 
to $F_M$. This requires a calculation of the matrix element 
$<\pi^+\pi^-|\overline{d}\sigma_{\mu\nu}(1,\gamma_5)s|K>$. 
Here we use naive dimensional analysis \cite{wein} to match the 
operator into a meson operator of the form
\begin{equation}
{\cal O}_M = g {e G_F\lambda \over  \sqrt{2}}  
\epsilon^{\mu\nu\alpha\beta}\partial_\mu \pi^+ 
\partial_\nu \pi^- F_{\alpha\beta}K
\label{mesonop}
\end{equation}
Following Weinberg \cite{wein} we obtain the order of magnitude 
estimate \cite{cheng}
\begin{equation}
g= {4\pi \over \sqrt{2}\lambda} {c \over \Lambda^2}
\end{equation}

There are many possible chiral Lagrangians that have the same 
transformation properties as the short distance operator, Eq.~\ref{sdgver}, 
that give rise to a meson operator like the one in Eq.~\ref{mesonop}. 
For example, guided by dimensional analysis, we can write down 
at ${\cal O}(p^4)$ the following effective Lagrangian
\begin{equation}
{\cal L} = g{e G_F \over \sqrt{2}}
\lambda f_\pi^3 \epsilon^{\mu\nu\alpha\beta}{\rm Tr}\biggl(
h \partial_\mu \Sigma \partial_\nu \Sigma^\dagger
\bigl(\Sigma F_{R\alpha\beta} + F_{L\alpha\beta}\Sigma\bigr)\biggr) 
\label{chilam}
\end{equation}
where the matrix $h$ is a three by three matrix with $h_{23}=1$ 
and all other elements zero to accomplish the $\Delta s=1$ transition. 
Notice that Eq.~\ref{chilam} has the same transformation properties 
as the short distance operator and is not suppressed by light-quark 
masses as is appropriate for the new physics interactions of 
interest, Eq.~\ref{cdef}. Within the standard model, the short 
distance operator is suppressed by light-quark masses and this 
results in a different effective chiral Lagrangian involving the usual 
chiral symmetry breaking factor. However, the standard model short 
distance coefficient is small enough to completely neglect its 
contribution. We have also included a Levi-Civita tensor to select 
the contribution of new physics to the  magnetic transition. For the 
radiative decays in question we replace $F_{R\mu\nu}=
F_{L\mu\nu}=eQ F_{\mu\nu}$ where  $Q$ is the quark charge matrix and 
$F_{\mu\nu}$ the photon field strength tensor.

After using the chiral Lagrangian, Eq.~\ref{chilam}, to calculate the 
amplitude for $K_L \rightarrow \pi^+ \pi^- \gamma$, we find 
\begin{equation}
\biggl|F_M\biggr|_{\rm New} = {\pi \over 2\sqrt{2}\lambda}{c \over f_\pi}
\end{equation}
which, from Eq.~\ref{fmbound} implies
\begin{equation}
|c(\mu)| \lsim 18~{\rm MeV}.
\end{equation}
This is comparable to the bound obtained from radiative hyperon decay, 
and can be improved with an improved determination of $F_M$. 

It is easy to check that Eq.~\ref{chilam} gives a similar 
contribution to the decay $K^+ \rightarrow \pi^+ \pi^0 \gamma$, 
but as argued before, this does not place additional constraints. 
Similarly, Eq.~\ref{chilam} does not contribute to 
$K_L \rightarrow \pi^0 \pi^0 \gamma$ in accordance with the fact 
that this amplitude starts at order $p^6$.

\section{Right handed $W$ couplings}

Left-right mixing is the obvious case in which the light-quark 
mass suppression of the operator Eq.~\ref{sdgver} can be removed, 
and turned into an enhancement from a heavy top-quark mass. Left-right 
symmetric models have been studied in the context of $b\to s\gamma$ \
in detail \cite{chomi}. For our purpose it suffices to look 
at the effective Lagrangian at the $W$ mass scale that results 
after integrating out a heavy right handed $W$. This can be 
done easily following the formalism of Peccei and Zhang \cite{pz}. 
In unitary gauge the new coupling of interest is
\begin{equation}
{\cal L}_{\rm eff} = {g\over \sqrt{2}} V_{UD}
\kappa^{UD}_R {\overline U}_R \gamma^\mu D_R W_\mu^+ + {\rm ~h.c.}
\label{ferml}
\end{equation}
In writing Eq.~\ref{ferml} we have assumed CP conservation and 
ignored modifications to the left-handed $W$ couplings which do 
not lead to enhanced effects. In general $\kappa^{UD}_R$ will be 
different for each $UDW$ coupling. 

This interaction has been considered before in the study of $b
\rightarrow s \gamma$ \cite{fuji}, and also in the study of CP 
violation in B decays \cite{alaa}. A trivial generalization of those 
results leads to:
\begin{eqnarray}
c_{11}^i&=&- {m_i \over m_s}\kappa^{is}_R F_{GR}(x_i),\nonumber\\
c_{12}^i&=&- {m_i \over m_s}\kappa^{is}_R F_{AR}(x_i),
\end{eqnarray}
where $i = u,c,t$,  $x_i=(m_i/m_W)^2$ and
\begin{eqnarray}
F_{GR}(x_i) &=& {6x_i\over (1-x_i)^3}\log(x_i)+{3(1+x_i)\over 
(1-x_i)^2}+1 \nonumber \\
F_{AR}(x_i) &=& {x_i (2-3x_i) \over (1-x_i)^3}\log(x_i)+
{5x_i^2-31x_i+20\over 6(1-x_i)^2}
\end{eqnarray}

In this case, the intermediate charm and top quark states lead to 
a value of $c$ much larger than in the standard model. 
If we assume that the $V_{UD}$ matrix elements of Eq.~\ref{ferml} 
are the same as the CKM elements in the SM, and use the approximate 
Wolfenstein parameterization for the CKM angles with $\rho=0$, 
we find at the $W$ mass scale (in units of MeV)
\begin{equation}
c(m_W) \approx -10 \kappa^{cs}_R - 2\kappa^{ts}_R.
\end{equation}
These numbers are reduced by about a factor of 2 at $\mu=1$~GeV 
when we use Eq.~\ref{running}. We find 
$c(\mu) \approx -6 \kappa^{cs}_R - 1.4\kappa^{ts}_R$. 

If we assume that 
$\kappa^{cs}_R$, $\kappa^{ts}_R$ are of the same order of magnitude,
then the charm-quark contribution is dominant. Ref.~\cite{fuji} has found 
that $b \to s \gamma$ constrains $\kappa^{tb}_R$ to be at most a 
few percent. Our strongest constraint from the radiative hyperon decays, 
$|c(\mu)| < 12$~MeV, implies that
\begin{eqnarray}
|\kappa^{cs}_R|  &\lsim & 2 \nonumber \\
|\kappa^{ts}_R|&\lsim & 8.5
\label{lrlimits}
\end{eqnarray}
which are not as restrictive as $b\rightarrow s \gamma$ if one assumes 
that all the $\kappa_R$ constants are of the same order. In the 
most general case our result constrains additional 
parameters in left-right symmetric models \cite{chomi}.


\section{Supersymmetric Models}

Another class of models in which the coefficient $c$ is naturally 
large is the general SUSY extension of the standard model. In this 
class of models one can generate the operator at one-loop via intermediate 
squarks and gluinos. The enhancement is due both to the strong 
coupling constant and to the removal of the chirality suppression 
that results in a gluino mass replacing the light-quark mass 
in Eq.~\ref{sdgver} \cite{Bertolini}. 

In the interaction basis, the down-quark-squark-gluino vertex 
is given by
\begin{equation}
{\cal L}=-\sqrt{2}g_{s}
(\overline{d}_{L}^{i}T^{a}\widetilde{g}_{a}\cdot \widetilde{D}_{L}^{i}
-\overline{d}_{R}^{i}T^{a}\widetilde{g}_{a}\cdot \widetilde{D}_{R}^{i})
\end{equation}
where $i$ is the generation index. Soft SUSY breaking squark masses will in
general induce mixing between different generations of squarks. 
The interaction eigenstates $\widetilde{D}_{L,R}^{i}$ are then different
from the mass eigenstates $\widetilde{D}^{k}$ inducing flavor changing 
neutral currents. Here we stay away from specific models and follow 
Ref.~\cite{ggms} to write the contributions to the 
$s\to d\gamma$ transition in the mass insertion approximation. 
The result found in Ref.~\cite{ggms}, after introducing an overall 
normalization to match our definition of $c_{11,12}$ in Eq.~\ref{sdgver}, 
is,
\begin{eqnarray}
&&c_{11} ={\sqrt{2}\alpha_s \pi \over G_F m^2_{\tilde q} V_{td}^*V_{ts}}
\biggl[\delta_{12}^{LL}\biggl({1\over 3} 
M_3(m^2_{\tilde g}/m^2_{\tilde q})
 +3 M_4(m^2_{\tilde g}/m^2_{\tilde q})\biggr)\nonumber\\
&&\;\;\;\;\;\;
+ \delta_{12}^{LR} {m_{\tilde g}\over m_s} 
\biggl({1\over 3}M_1(m^2_{\tilde g}/m^2_{\tilde q}) 
+ 3 M_2(m^2_{\tilde g}/m^2_{\tilde q})\biggr)\biggr]
\nonumber \\
&&c_{12} ={\sqrt{2}\alpha_s \pi \over G_F m^2_{\tilde q} V_{td}^*V_{ts}}
{8\over 9}\biggl[\delta_{12}^{LL} 
M_3(m^2_{\tilde g}/m^2_{\tilde q})
+ \delta_{12}^{LR} {m_{\tilde g}\over m_s} 
M_1(m^2_{\tilde g}/m^2_{\tilde q})\biggr];\nonumber\\
\end{eqnarray}
where the loop-functions are given by
\begin{eqnarray}
&&M_1(x)={1+4x-5x^2+2x(2+x)\ln x\over 2(1-x)^4};\nonumber\\
&&M_2(x)=-{5-4x-x^2+2(1+2x)\ln x\over 2(1-x)^4};\nonumber\\
&&M_3(x)= {-1+9x+9x^2-17x^3+6x^2(3+x)\ln x\over 12(x-1)^5};\nonumber\\
&&M_4(x)= {-1-9x+9x^2+x^3-6x(1+x)\ln x\over 6(x-1)^5}.
\end{eqnarray}
The parameters $\delta_{12}^{}$ characterize the mixing in the mass insertion
approximation \cite{ggms}, $x=m_{\tilde g}^2/m_{\tilde q}^2$, 
and $m_{\tilde g}$, $m_{\tilde q}$ are the average gluino and 
squark masses, respectively. In general  
the quark-squark-gluino interactions also generate the $s\to d \gamma$
transition with different chiralities. This can be obtained by
replacing $\delta_{12}^{LL}$ and $\delta_{12}^{LR}$ by $\delta_{12}^{RR}$
and $\delta_{12}^{RL}$ in the above expressions. 

As anticipated, there is an enhancement factor $m_{\tilde g}/m_s$ in 
the term proportional to $\delta_{12}^{LR}$ and this is the term with a 
potentially large contribution to $s\to d \gamma$.

In Table~\ref{tsusy} we illustrate some representative values 
for $c$ both at a scale $\mu=1$~GeV and at the $W$ mass scale. 
The real part of $\delta_{12}^{LR}$ is constrained from 
$\Delta m_{K_S-K_L}$ to be typically less than a few times 
$10^{-2}$ \cite{ggms}. This is to be compared with our 
constraint for $x=0.3$, 
\begin{equation}
\delta_{12}^{LR} \lsim 0.14
\end{equation}
The imaginary part is more severely
constrained from $\epsilon'/\epsilon$\cite{ggms}. 
Even though our constraint is not as good as that from 
$K$-$\overline{K}$ mixing the two are complementary in 
a general model. Our bound would be important, for example, 
in models that try to avoid the $K$-$\overline{K}$ mixing 
bound through an interplay of $\delta_{12}^{LR}$ and 
$\delta_{12}^{RL}$.

\begin{table}[htb]
\centering
\begin{tabular}{|c|c|c|} \hline
x& $c(\mu)$&$c(m_W)$\\ \hline
&&\\
0.3&$88\,\delta_{12}^{LR}$&$73\,\delta_{12}^{LR}$\\
1&$57\,\delta_{12}^{LR}$&$60\,\delta_{12}^{LR}$\\
4&$24\,\delta_{12}^{LR}$&$32\,\delta_{12}^{LR}$\\
& & \\ \hline
\end{tabular}
\caption{ $c$ (in units of MeV) in generalized SUSY models 
due to $\delta_{12}^{LR}$ with $m_{\tilde q} = 500$ GeV 
for different values of $x=m^2_{\tilde g}/m^2_{\tilde q}$.}
\label{tsusy}
\end{table}

\section{Conclusions}

We have studied radiative hyperon decays and radiative 
kaon decays of the form $K \to \pi \pi \gamma$ with the 
aim of constraining new interactions. Guided by chiral 
perturbation theory and dimensional analysis we have 
parameterized the long distance physics that dominates these 
modes and in this way we have quantified a bound on possible 
contributions from short-distance new physics. We found that 
these processes can place interesting constraints on certain 
models in which the $s\to d \gamma$ transition is significantly 
enhanced with respect to the standard model. 

\vspace{1in}

\noindent {\bf Acknowledgments} The work of G.V. was supported in
part by DOE under contract number DEFG0292ER40730. 
The work of X-G.H. was supported by NSC of R.O.C. under grant number
NSC88-2112-M-002-041 and Australian Research Council. 
We thank Bruce McKellar and the theory group at Melbourne 
for their hospitality while part of this work was completed. 
We also thank F.~Gabbiani for discussions on Ref.~\cite{ggms} and  
A.~Kagan for bringing Ref.~\cite{kagan} to our attention.

\end{document}